# Flow through an Electromechanical Tube Driven by a Voltage

Yilin Qu[1,2], Hendrik J. Viljoen[3], Jiashi Yang[4,a]

[1]School of Marine Science and Technology, Northwestern Polytechnical University, Xi'an, Shaanxi 710072, China

[2]Unmanned Vehicle Innovation Center, Ningbo Institute of NPU, Ningbo, Zhejiang 315048, China

[3]Department of Chemical and Biomolecular Engineering, University of Nebraska-Lincoln, Lincoln, NE 68588, USA

[4]Department of Mechanical and Materials Engineering, University of Nebraska-Lincoln, Lincoln, NE 68588, USA

a) Corresponding author: jyang1@unl.edu

*Abstract* – We study the axial flow of a nonviscous and incompressible fluid in a circular tube made from an electromechanical material. The tube is driven into radial motion by an electric voltage across its thickness. A theoretical analysis is performed. An analytical solution is obtained from the relevant equations of fluid mechanics and piezoelectric shells. The solution shows that when the tube is contracting or expanding radially under a properly designed voltage, the fluid can flow along the tube axially. Hence the tube can function as a pump for driving the fluid. The effects of various physical parameters on the velocity profile of the flow are examined.

## 1. Introduction

Electromechanical materials possess piezoelectric [1,2], electrostrictive [3,4] or flexoelectric [5-7] couplings. They deform under an electric field and become electrically polarized under stress, strain or strain gradients. Electromechanical materials are widely used to make devices for various applications such as transducers for electromechanical energy conversion [8-10], physical and chemical sensors [11-14], actuators in smart structures [15], ultrasonic motors [16,17], gyroscopes for angular rate sensing [18-20], acoustic wave resonators [21,22] and filters [23] for frequency selection and operation, energy harvesters [24-27], transformers for raising an electric voltage [28,29], and through-wall power delivery [30,31].

In particular, piezoelectric materials can be used to make fluid pumps [32-34]. Piezoelectric pumps typically have complicated structures with piezoelectric and nonpiezoelectric components as well as chambers and valves, etc. for which theoretical analyses are rare [32]. In this paper we explore the possibility of using an electromechanical tube for fluid pumping. A theoretical analysis is performed using the equations of piezoelectricity and the equations of fluid mechanics. Through a one-dimensional analysis of the axial motion of the fluid, it is shown that an axial flow of the fluid can be produced when a proper electrical voltage is applied to the tube.

## 2. Governing Equations

Consider a nonviscous fluid whose velocity field **v**(**x**,*t*) and pressure field *p*(**x**,*t*) are governed by Euler's equation [35]:

$$\frac{\partial \mathbf{v}}{\partial t} + (\mathbf{v} \cdot \nabla)\mathbf{v} = -\frac{1}{\rho}\nabla p, \qquad (1)$$

where $\rho$ is the mass density of the fluid. $\rho$ is assumed to be a constant for an incompressible fluid which we are considering. We also limit ourselves to irrotational motions. Hence

$$\nabla \cdot \mathbf{v} = 0, \qquad (2)$$
$$\nabla \times \mathbf{v} = 0. \qquad (3)$$

(3) allows the introduction of a velocity potential $\phi$ such that
$$\mathbf{v} = \nabla \phi. \tag{4}$$
Then (3) is no longer needed and (2) takes the following form:
$$\nabla^2 \phi = \frac{\partial^2 \phi}{\partial x_1^2} + \frac{\partial^2 \phi}{\partial x_2^2} + \frac{\partial^2 \phi}{\partial x_3^2} = 0. \tag{5}$$
With (4), we can integrate (1) to obtain
$$\frac{\partial \phi}{\partial t} + \frac{1}{2} v^2 + \frac{p}{\rho} = f(t), \tag{6}$$
where $f(t)$ is an arbitrary function of time and
$$v^2 = \mathbf{v} \cdot \mathbf{v}. \tag{7}$$

The electromechanical tube under consideration is shown in Fig. 1. Its average radius is $R$ and its thickness is $2h$. We assume a thin tube with $R \gg 2h$. The tube has many pairs of circular ring electrodes on its inner and outer surfaces so that different voltages can be applied to different pairs of inner and outer electrodes at any locations and any time. In the special case when the tube is made from polarized ceramics such as PZT-4, the poling direction is along the thickness of the wall of the tube as shown.

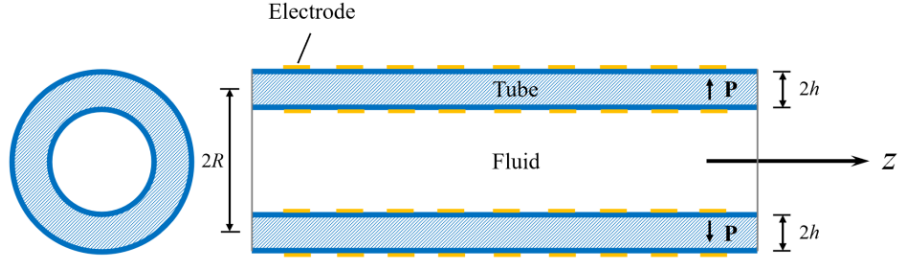

Fig. 1. An incompressible fluid in a circular electromechanical tube with driving electrodes.

The tube can be modeled as a circular cylindrical shell in axisymmetric motion with a radial displacement $u_r(z,t)$ in cylindrical coordinates. In the case of polarized ceramics, from the theory of piezoelectric shells [36], $u_r$ is governed by
$$-N_\theta \frac{1}{R} + p\big|_{r=R} = 2\rho' h \frac{\partial^2 u_r}{\partial t^2}, \tag{8}$$
where $\rho'$ is the mass density of the shell. $N_\theta$ is the circumferential extensional force in the shell. $N_\theta/2h$ is the so-called hoop stress in the shell. $N_\theta$ is related to the radial displacement $u_r$ and the radial electric field $E_r$ produced by the applied voltage through [36]
$$N_\theta = 2h\left(c_{11}^p \frac{u_r}{R} - e_{31}^p E_r\right), \tag{9}$$
where the effective elastic constant $c_{11}^p$ and effective piezoelectric constant $e_{31}^p$ for thin shells are related to the usual three-dimensional material constants by [36]
$$\begin{aligned} c_{11}^p &= c_{11} - c_{13}^2 / c_{33}, \\ e_{31}^p &= e_{31} - c_{13} e_{33} / c_{33}. \end{aligned} \tag{10}$$
Substituting (9) into (8), we obtain the following equation for the radial motion of the tube:
$$-\left(c_{11}^p \frac{u_r}{R} - e_{31}^p E_r\right)\frac{2h}{R} + p\big|_{r=R} = 2\rho' h \frac{\partial^2 u_r}{\partial t^2}. \tag{11}$$

## 3. Determination of Axial Flow

For axisymmetric motions of the fluid, in cylindrical coordinates, we have

$$v_r = \frac{\partial \phi}{\partial r}, \quad v_\theta = 0, \quad v_z = \frac{\partial \phi}{\partial z}, \tag{12}$$

$$\nabla \cdot \mathbf{v} = \frac{1}{r}\frac{\partial}{\partial r}(rv_r) + \frac{\partial v_z}{\partial z} = 0. \tag{13}$$

We integrate (13) over a cross section of the tube which leads to

$$\begin{aligned}
\int_0^R 2\pi r \left[\frac{1}{r}\frac{\partial}{\partial r}(rv_r) + \frac{\partial v_z}{\partial z}\right] dr \\
= 2\pi \int_0^R \frac{\partial}{\partial r}(rv_r) dr + \int_0^R 2\pi r \frac{\partial v_z}{\partial z} dr \\
= 2\pi [rv_r]_0^R + \frac{\partial}{\partial z}\int_0^R 2\pi r v_z dr \\
= 2\pi R v_r|_R + \frac{\partial Q_z}{\partial z} = 2\pi R v_r|_R + \pi R^2 \frac{\partial \bar{v}_z}{\partial z} = 0,
\end{aligned} \tag{14}$$

where the axial flux $Q_z$ of the flow and the average axial velocity $\bar{v}_z$ are defined by

$$Q_z = \int_0^R 2\pi r v_z dr = \pi R^2 \bar{v}_z. \tag{15}$$

At the inner surface of the tube, the continuity of the normal velocity between the fluid and the tube is

$$v_r|_R = \frac{\partial u_r}{\partial t}. \tag{16}$$

From (11) we have

$$\begin{aligned}
2\pi R \frac{\partial u_r}{\partial t} = \frac{2\pi R^2 e_{31}^p}{c_{11}^p} \frac{\partial E_r}{\partial t} + \frac{\pi R^3}{h c_{11}^p} \frac{\partial}{\partial t}\left(p|_{r=R}\right) \\
- \frac{2\pi R^3 \rho'}{c_{11}^p} \frac{\partial^3 u_r}{\partial t^3} = 2\pi R v_r|_R,
\end{aligned} \tag{17}$$

where (16) has been used. The substitution of (17) into (14) yields

$$\begin{aligned}
\frac{\partial Q_z}{\partial z} = -\frac{2\pi R^2 e_{31}^p}{c_{11}^p} \frac{\partial E_r}{\partial t} \\
- \frac{\pi R^3}{h c_{11}^p} \frac{\partial}{\partial t}\left(p|_{r=R}\right) + \frac{2\pi R^3 \rho'}{c_{11}^p} \frac{\partial^3 u_r}{\partial t^3},
\end{aligned} \tag{18}$$

or

$$\frac{\partial \bar{v}_z}{\partial z} = -\frac{2 e_{31}^p}{c_{11}^p} \frac{\partial E_r}{\partial t} - \frac{R}{h c_{11}^p} \frac{\partial}{\partial t}\left(p|_{r=R}\right) + \frac{2R\rho'}{c_{11}^p} \frac{\partial^3 u_r}{\partial t^3}. \tag{19}$$

The first term on the right-hand side of (19) is the direct effect of $E_r$ on $\bar{v}_z$. The second term describes the interaction between the tube and the fluid. The third term is due to the inertia of the tube. To focus on the most basic effect of $E_r$ on the fluid, we drop the second and the third terms on the right-hand side of (19). This approximation is equivalent to allowing the tube to expand or contract freely and instantly under $E_r$, with $u_r$ determined from the following approximate form of (11):

$$-\left(c_{11}^p \frac{u_r}{R} - e_{31}^p E_r\right)\frac{2h}{R} + 0 = 0. \tag{20}$$

In this case (19) reduces to

$$\frac{\partial \bar{v}_z}{\partial z} = -\frac{2e_{31}^p}{c_{11}^p}\frac{\partial E_r}{\partial t}. \tag{21}$$

(21) shows that the axial flow is proportional to the piezoelectric constant of the tube and inversely proportional to its stiffness, which is reasonable. From (21) we can determine the axial velocity $\bar{v}_z$ from the applied $E_r$.

Consider the case when the driving electric field is given by

$$E_r = F(z - ct), \tag{22}$$

where $F$ is an arbitrary function and $c$ is a constant. The electric field described by (22) is like a wave propagating to the right of the tube with a speed $c$. Then, from (21),

$$\frac{\partial \bar{v}_z}{\partial z} = c\frac{2e_{31}^p}{c_{11}^p}F', \tag{23}$$

where

$$F' = \frac{dF}{d(z-ct)}. \tag{24}$$

Integrating (23) with respect to $z$, we obtain

$$\bar{v}_z = c\frac{2e_{31}^p}{c_{11}^p}F(z-ct) + B(t). \tag{25}$$

where $B$ is an arbitrary function of time. Consider an infinite tube with a given fluid velocity $v_0$ at its left end. Then

$$\bar{v}_z\big|_{z=-\infty} = v_0(t). \tag{26}$$

In this case, from (25) and (26), we obtain

$$B = -c\frac{2e_{31}^p}{c_{11}^p}F(-\infty) + v_0. \tag{27}$$

Hence,

$$\bar{v}_z = c\frac{2e_{31}^p}{c_{11}^p}\left[F(z-ct) - F(-\infty)\right] + v_0. \tag{28}$$

### 4. Numerical Results and Discussion

As a specific example, consider

$$E_r = F = E\left[\frac{\pi}{2} - \tan^{-1}(z-ct)\right], \tag{29}$$

$$F(-\infty) = \pi,$$

where $E$ and $c$ are constants. $E$ represents the amplitude of the applied electric field. Then, from (28), we obtain the axial velocity as

$$\bar{v}_z = c\frac{2e_{31}^p}{c_{11}^p}E\left[-\frac{\pi}{2} - \tan^{-1}(z-ct)\right] + v_0. \tag{30}$$

For the material of the tube, we use PZT-4 as an example whose material constants are

$$c_{11} = 139\text{GPa}, \quad c_{13} = 74\text{GPa}, \quad c_{33} = 115\text{GPa},$$

$$e_{31} = -5.2 \text{ C/m}^2, \quad e_{33} = 15.1 \text{ C/m}^2.$$

The radius of the tube is taken to be $R = 1$ cm. The amplitude of the applied electric field is $E = 10^6$ N/C. $c$ in the applied electric field is 1 cm/s. The axial velocity of the fluid at $z = -\infty$ is $v_0 = 0$.

Some of these parameters will be varied one at a time below. With these parameters, the amplitude of the radial displacement of the shell is found to be

$$u_r = R \frac{e^p_{31}}{c^p_{11}} E_r(-\infty) = \pi R \frac{e^p_{31}}{c^p_{11}} E \quad (31)$$

$$= -5.1281 \times 10^{-4} \text{ cm}.$$

Figure 2 shows the deformed tube under the applied electric field and the distribution of the axial flow at a few time instants qualitatively. Under $E$, the left part of the tube contracts according to (20). This contraction propagates to the right and the fluid flows to the right accordingly. Hence the tube can function as a pump for driving the fluid axially when an electric field is applied properly.

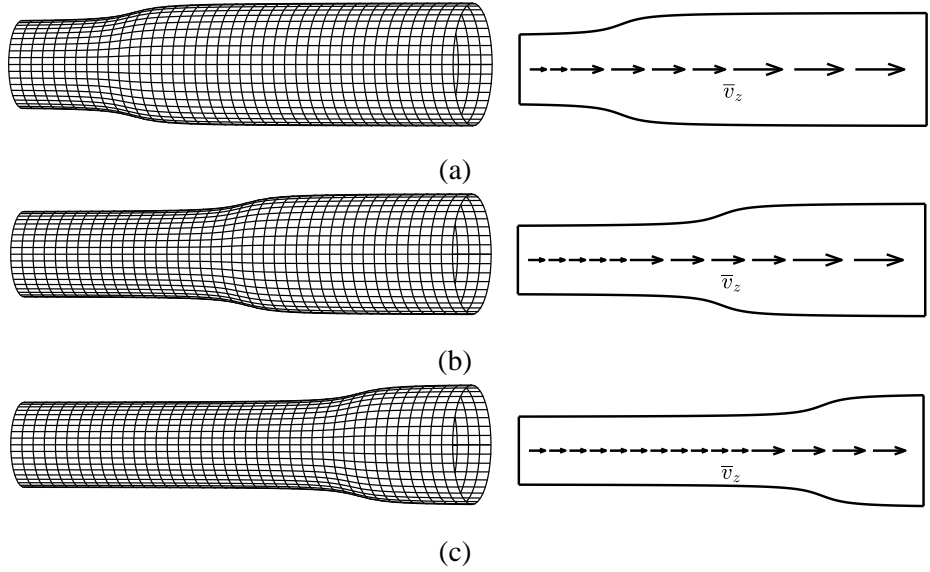

Fig. 2. Deformed tube and axial flow. (a) $t = 0$ s. (b) $t = 5$ s. (c) $t = 10$ s.

Figure 3 shows quantitatively the applied electric field and the axial velocity distribution for different time instants. As the wave-like electric field propagates to the right, the velocity distribution also propagates to the right.

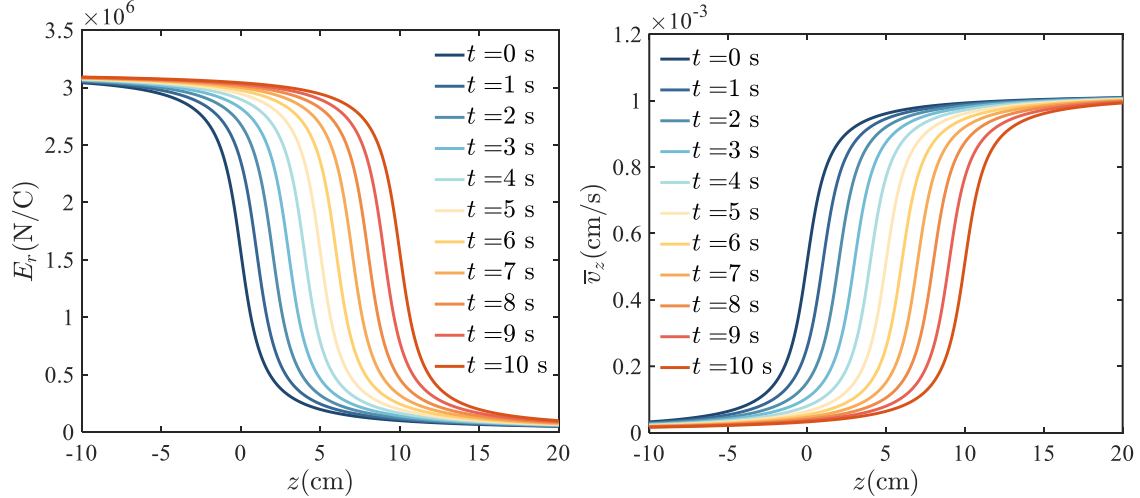

Fig. 3. Applied electric field (a) and axial velocity (b) at different time instants.

Figure 4 shows the effect of $c$ in (29) and (30). $c$ may be viewed as the speed at which the applied electric field moves to the right. For larger values of $c$, the electric field moves faster to the right and so does the axial velocity distribution. In addition, as indicated by (30), the amplitude of the axial velocity increases with $c$, which is reasonable.

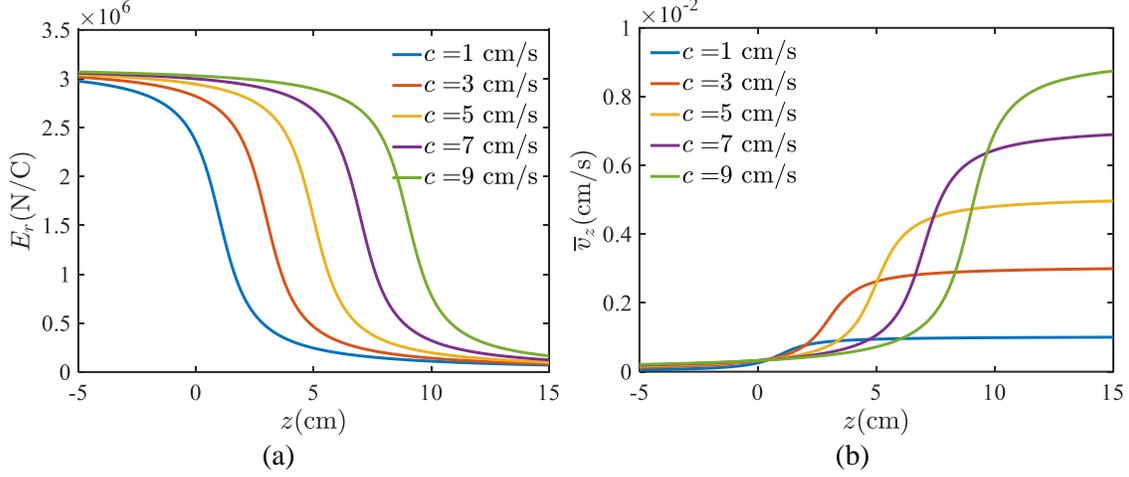

(a)          (b)

Fig. 4. Effect of $c$ on the applied electric field (a) and axial velocity (b). $t = 1$ s.

Figure 5 shows the effect of incoming velocity $v_0$ at the left end of the tube which may be positive or negative. Since the left part of the tube is in contraction under a positive $E$, an increase of $v_0$ contributes to the flow of the fluid to the right as shown.

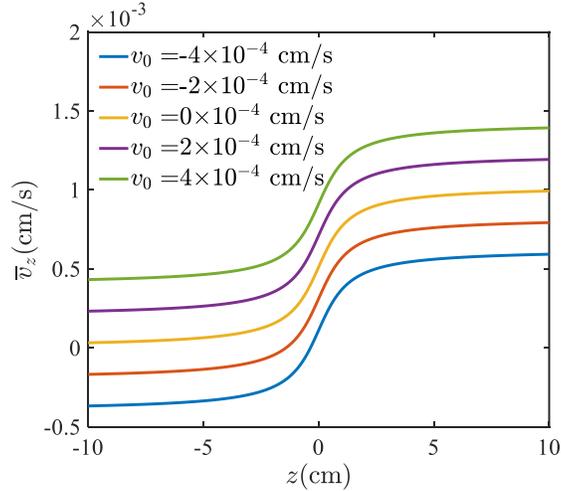

Fig. 5. Effect of $v_0$ on the distribution of the axial velocity. $t = 0$ s.

Figure 6 shows the effect of the amplitude $E$ of the driving electric field. When $E$ is positive and increasing, the electric field becomes stronger, the left part of the tube contracts more, and the axial velocity to the right becomes larger accordingly. If the applied electric field is reversed by changing its sign, the left part of the tube expands and the fluid flows to the left along the tube as shown in Fig. 7 which is the opposite of Fig. 2.

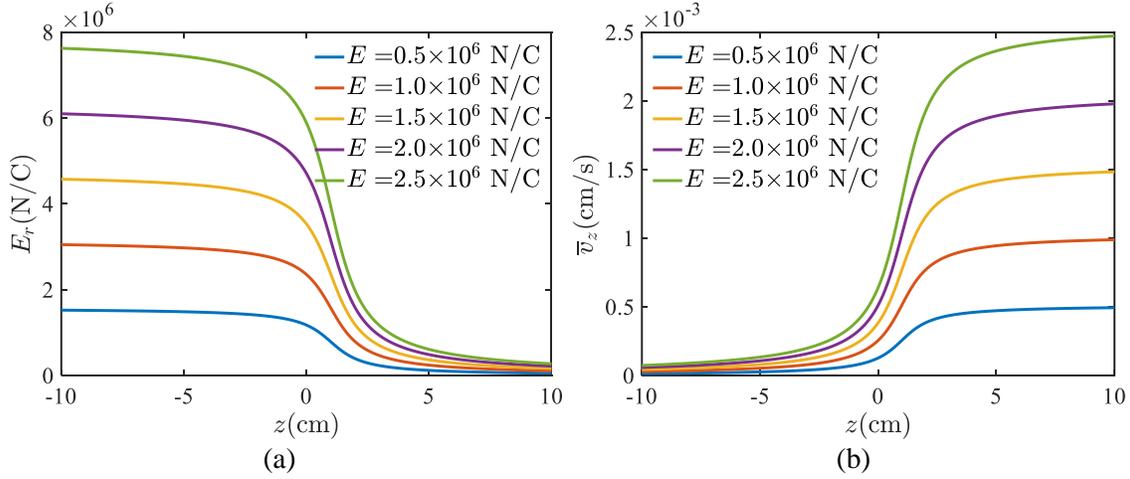

Fig. 6. Effect of *E* on the applied electric field (a) and axial velocity (b). $t = 0$ s.

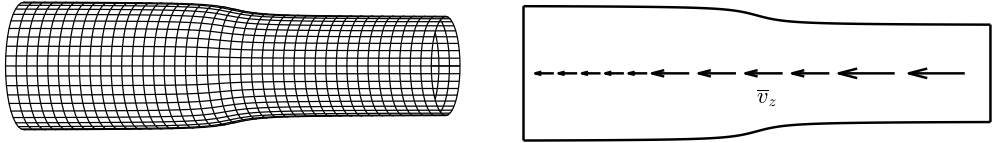

Fig. 7. Deformed tube and axial flow. $t = 0$ s. $E = -10^6$ N/C.

## 5. Conclusions

An electromechanical tube can operate as a fluid pump. Under a properly designed driving voltage, an axial flow of the fluid inside the tube can be created. The axial flow velocity is proportional to the effective piezoelectric constant of the tube and is inversely proportional to its effective elastic constant. If the tube is made from soft electromechanical materials [37] with large deformations, the axial flow is expected to be significantly stronger.

## Acknowledgements

The first author was supported by the China Postdoctoral Science Foundation (Grant No. 2023M732863).